\documentclass[10pt,conference]{IEEEtran}
\IEEEoverridecommandlockouts

\usepackage{cite}
\usepackage{amsmath,amssymb,amsfonts}
\usepackage{graphicx}
\usepackage{textcomp}
\usepackage{xcolor}
\usepackage{url}
\usepackage{algorithm}
\usepackage{algorithmic}
\usepackage{booktabs}
\usepackage{multirow}

\def\BibTeX{{\rm B\kern-.05em{\sc i\kern-.025em b}\kern-.08em
    T\kern-.1667em\lower.7ex\hbox{E}\kern-.125emX}}
\begin{document}

\title{WarmTuner: Program-Specific Warm Starts for Compiler Autotuning via Offline-to-Online Reinforcement Learning}

\author{\IEEEauthorblockN{1\textsuperscript{st} Tianlu Qiao}
\IEEEauthorblockA{
\textit{School of Electronic and Computer Engineering}\\
\textit{Peking University Shenzhen Graduate School}\\
Shenzhen, China\\
tlqiao25@stu.pku.edu.cn}
\and
\IEEEauthorblockN{2\textsuperscript{nd} Mingxuan Zhu}
\IEEEauthorblockA{\textit{Key Lab of High Confidence Software Technologies}\\
\textit{(Peking University), Ministry of Education}\\
\textit{Peking University}\\
Beijing, China \\
zhumingxuan@stu.pku.edu.cn}
\and
\IEEEauthorblockN{3\textsuperscript{rd} Zeyu Sun}
\IEEEauthorblockA{\textit{National Key Laboratory of Space Integrated Information System} \\
\textit{Institute of Software, Chinese Academy of Sciences}\\
Beijing, China \\
zeyu.zys@gmail.com}
\and
\IEEEauthorblockN{4\textsuperscript{th} Dan Hao\textsuperscript{*}}
\IEEEauthorblockA{
\textit{School of Electronic and Computer Engineering}\\
\textit{Peking University Shenzhen Graduate School}\\
Shenzhen, China\\
haodan@pku.edu.cn}
\thanks{\textsuperscript{*}Dan Hao is the corresponding author.}
}
\def\methodName#1{WarmTuner{#1}}
\newcommand{\methodVariant}[1]{\methodName{$_{\mathit{#1}}$}}
\newcommand{\methodNoPre}{\methodVariant{NoPre}}
\newcommand{\methodNoRep}{\methodVariant{NoRep}}
\newcommand{\methodReinf}{\methodVariant{Reinf}}
\newcommand{\methodPPO}{\methodVariant{PPO}}
\newcommand{\methodPDCAT}{\methodVariant{PDCAT}}
\newcommand{\methodKFour}{\methodName{$_{K=4}$}}
\newcommand{\methodKSixteen}{\methodName{$_{K=16}$}}
\maketitle

\begin{abstract}

Compilers are fundamental software tools that translate high-level programs into machine code. Modern compilers expose hundreds of optimizations, each turned on or off through an optimization flag, to improve the performance of the generated code. 
However, the number of possible flag combinations grows exponentially, making it difficult to find a flag configuration well suited to a given target program.
Existing compiler auto-tuning techniques reduce tuning cost by pruning the search space, injecting search biases, or predicting configuration performance. Although some exploit program features, the knowledge they extract from historical data is frozen once search begins; runtime feedback then guides only the search itself, never the prior. As a result, when this prior mismatches the target program, these methods waste much of the limited online budget before the search reaches good configurations.
We propose \methodName{}, an offline-to-online reinforcement learning framework that instead turns historical records into a program-conditioned \emph{policy} that predicts each flag's setting over the full flag space and remains adaptable on the target program.
Offline, \methodName{} learns this program-conditioned policy over the full flag space from historical good configurations. Online, it refines the same policy on the target program using real compile-run feedback, so that the policy is driven by measured speedups rather than limited to the historical data. 
Because each reward is expensive, yet directly verifiable, we instantiate the online update with Group Relative Policy Optimization (GRPO), which derives its signal from the relative speedups of candidates in the same round and does not need a separate value model—a natural fit for compiler tuning.
To evaluate the performance of \methodName{}, we conducted an extensive experimental study using the latest version of the GCC compiler 15.2.0 on two widely used benchmarks, cBench and PolyBench. The results show that \methodName{} achieves an average speedup of $1.732\times$ over GCC -O3 and obtains the best result on 14/30 programs, significantly outperforming the compared techniques. 
\end{abstract}


\section{Introduction}

Compilers are a core part of modern software development, which transform programs written in high-level languages into executable machine code. During this process, compiler can apply compiler optimizations to achieve objectives such as reducing the size of the compiled code or improving its runtime performance. Although GCC provides default optimization settings such as \verb|-O1|, \verb|-O2|, and \verb|-O3|, these settings are designed for general use and do not always perform well on a specific program, possibly resulting in suboptimal runtime performance. It is therefore valuable to find a program-specific optimization setting, particularly for programs that run frequently or whose execution is time-consuming. This problem is commonly studied as compiler auto-tuning, which includes selecting which compiler optimizations to apply (i.e., \emph{flag selection}) and deciding the order in which they are applied (i.e., \emph{phase ordering})~\cite{compiler_autotuning_survey}. As phase ordering does not apply to most production compilers such as GCC, this paper focuses on flag selection.

Common compilers like GCC provide hundreds of optimizations (e.g., loop unrolling), each controlled by enabling or disabling a specific optimization flag (e.g., \verb|-funroll-loops|). Enabling an optimization flag does not always improve runtime performance, and its effect often depends on the other enabled flags and on the structure of the target program. Because of the large number of flags and their exponential combinations, it is difficult to understand the impact of each flag and to manually choose a good flag configuration for a given program. To reduce this human effort, numerous auto-tuning techniques that focus on flag selection have been proposed~\cite{agakov_iterative_optimization,almagor_effective_sequences,boca,garciarena_ga_flags,kisuki2002iterative,li2014feature,cfsca,multiphase_learning}.

Compiler auto-tuning is a typical search problem, i.e., searching a good flag configuration within the space of optimizations, and thus the existing techniques can be viewed as different ways of making the search problem easier. Traditional search-based techniques explore candidate configurations through hill climbing~\cite{almagor_effective_sequences}, genetic or evolutionary algorithms~\cite{cole,ga_gcc_options,garciarena_ga_flags}, random iterative optimization~\cite{rio}, automatic configuration tools such as irace~\cite{irace_gcc}, and multi-technique frameworks such as OpenTuner~\cite{opentuner}. Since each candidate configuration must be compiled and executed, learning-based techniques have been proposed to reduce this cost by modeling the performance of the configuration and guiding the search. For example, BOCA uses Bayesian optimization to select promising compiler configurations under a limited tuning budget~\cite{boca}, COBAYN builds Bayesian networks to model relationships among program features, optimization flags, and performance~\cite{cobayn}, and CompTuner learns performance prediction models through multiple phases and uses particle swarm optimization to search for effective flag combinations~\cite{multiphase_learning}. More recent work improves search efficiency by exploiting structure in the optimization space, which can be classified as structure-aware techniques: CFSCA identifies program-relevant critical flags and searches more intensively over them~\cite{cfsca}, PDCAT uses compiler-documentation constraints and flag preferences to reduce and bias the search space~\cite{pdcat}, SRTuner mines synergistic relations among flags~\cite{srtuner}, and GroupTuner performs localized mutation over coherent option groups~\cite{grouptuner}.

These techniques facilitate the searching process in different ways, but none of them address the critical problem in the searching process: \emph{where should the search start for a target program, and can that starting point still be corrected during search?} In particular, the search-based techniques~\cite{irace_gcc,opentuner} rely on generic or weakly informed starting points and may spend much of the online budget just reaching a useful region of the flag space. Learning-based techniques~\cite{boca,cobayn} predict promising configurations, but even a strong predictor rarely gives the best one for every program in a single shot. Structure-aware techniques~\cite{cfsca,srtuner} narrow the search space using mined or largely program-agnostic structure, yet still do not give each program its own starting distribution. The closest prior work to this question is PDCAT~\cite{pdcat}, which explicitly recognizes the importance of the initial search bias: it uses historical records to set initial flag-level probabilities and then uses this bias to guide subsequent search. However, this bias is still program-agnostic. Once the initial flag probabilities are determined, the same starting distribution is used for all target programs, regardless of their source-code characteristics. To sum up, existing priors are either generic, shared across programs, or fixed once online search begins. Runtime feedback may steer the search process under these priors, but it does not update a program-conditioned prior that generates a different starting distribution for each target program. As a result, programs with very different optimization preferences may still begin from similar regions of the flag space, wasting budget that a program-specific, adaptable policy could save.

Our key insight is twofold. First, historical tuning data, accumulated across many programs, reveal how program characteristics relate to promising regions of the flag space, and can therefore guide where the search should start for each particular program. Second, an offline starting point alone is not enough: a new target program may differ from those seen offline. Therefore, the tuner must keep adapting to the behavior observed on the target program, which naturally calls for online reinforcement learning driven by measured speedups over \verb|-O3|.

Based on these observations, we propose an Offline-to-online Reinforcement LeArning framework for compiler auto-tuning (abbreviated as \methodName{}). The offline stage is designed to provide a customized starting point for each target program. Instead of using a shared initial configuration, a manually designed preference bias, or a one-shot prediction of the final best configuration, \methodName{} learns a program-conditioned flag-selection policy from historical tuning records. Given the source-code representation of a target program and the \verb|-O3| reference state, the policy outputs flag-enable probabilities over the full flag space. These probabilities define a program-specific starting point, so different programs can begin online tuning from different regions that match their predicted optimization preferences. The online stage further refines this customized starting point with feedback from the same target program. \methodName{} samples candidate configurations from the program-specific policy, measures their speedups over \verb|-O3|, and updates the same policy using the measured rewards. Therefore, the initial distribution is not only customized offline, but also correctable online when it mismatches the target program. This differs from methods whose starting bias may guide subsequent search but remains external to, or fixed relative to, the policy that generates future configurations. 

We instantiate the online update with a reinforcement learning algorithm, Group Relative Policy Optimization (GRPO)~\cite{deepseekmath_grpo}. 
In each online round, \methodName{} samples a group of candidate flag configurations from the current policy, compiles and executes them on the target program, and uses their measured speedups over \verb|-O3| as rewards. 
The policy is then updated according to the relative performance of candidates in the same group, so that later rounds sample configurations from a policy already adapted to the target program. 
This process repeats until the tuning budget is exhausted, and \methodName{} returns the best measured flag configuration.

To evaluate \methodName{}, we conducted experiments on GCC 15.2.0 using two widely used compiler auto-tuning benchmarks, cBench~\cite{cbench} and PolyBench~\cite{polybench}. We compare \methodName{} with four representative compiler auto-tuning techniques, including RIO~\cite{rio}, SRTuner~\cite{srtuner}, GroupTuner~\cite{grouptuner}, and PDCAT~\cite{pdcat}. 
Under the 5,000s budget, the flag configurations generated by \methodName{} outperform the default \verb|-O3| setting on all 30 selected evaluation programs, with speedups ranging from $1.133\times$ to $4.159\times$ (i.e., 13.3\%--315.9\% improvement over \verb|-O3|). 
Among the four compared techniques, \methodName{} achieves the highest average speedup ($1.732\times$) and significantly outperforms RIO, SRTuner, and PDCAT. 
Against the strongest baseline, GroupTuner, \methodName{} still obtains a higher average speedup ($1.732\times$ vs. $1.669\times$), more best results (14 vs. 11), and fewer worst results (1 vs. 4). 
To study the impact of its components, we construct 7 variants of \methodName{}; the ablation results show that the main components improve tuning performance.

The contributions of this paper are summarized as follows:
\begin{enumerate}
    \item \textbf{An offline-to-online reinforcement learning framework for compiler auto-tuning}, which learns an initial flag selection policy from historical tuning data and improves it on the target program through online policy optimization with real execution feedback.

    \item \textbf{An extensive experiment} on the latest version of GCC (i.e., GCC 15.2.0) using two widely used benchmarks, cBench and PolyBench, demonstrates the performance of \methodName{} as well as its components.

    \item \textbf{A reproducible package}, available in a public repository~\cite{zenodo}.
\end{enumerate}

\begin{figure*}[htbp]
\centerline{\includegraphics[width=0.95\textwidth]{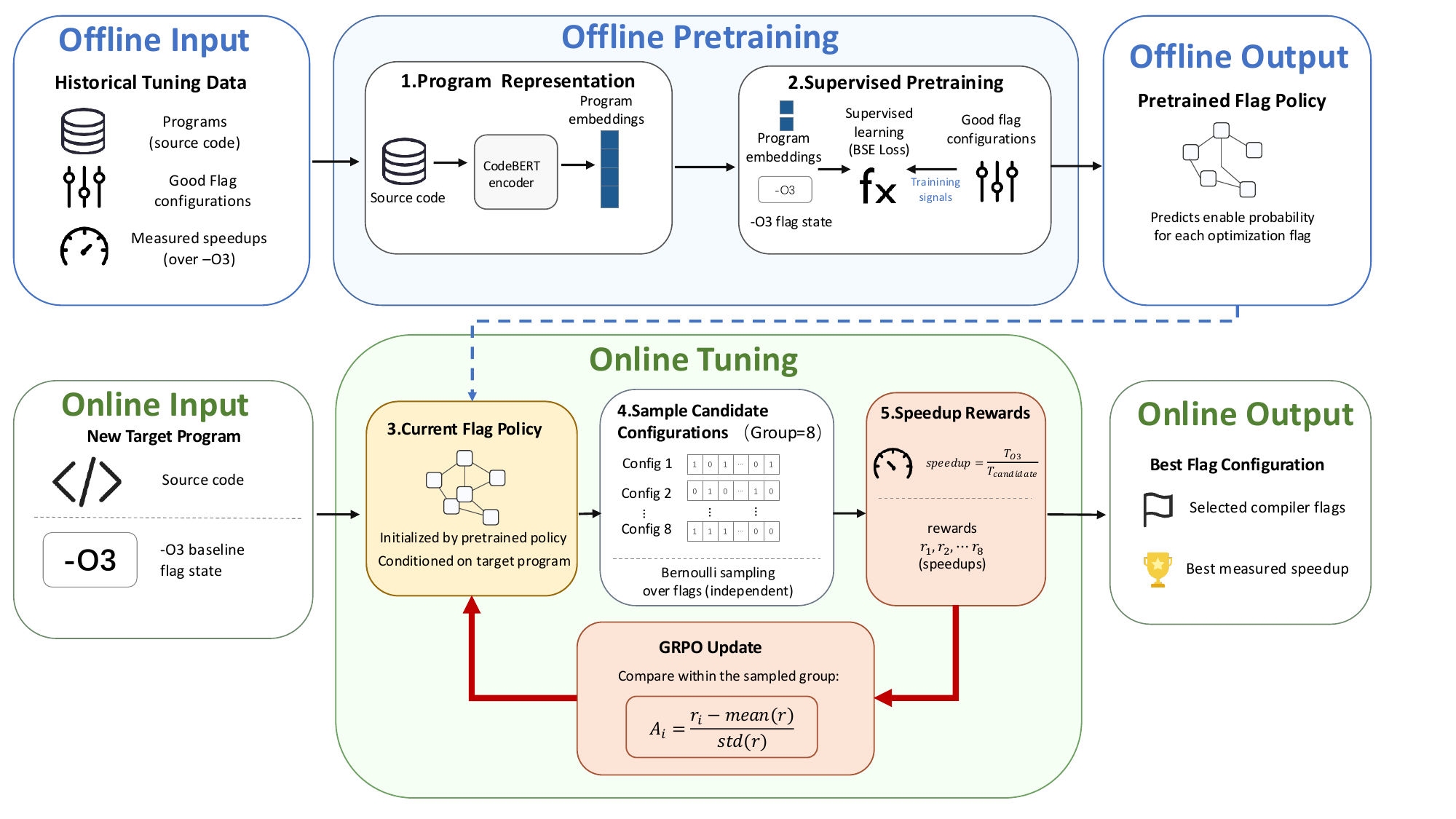}}
\caption{Overview of \methodName{}. }
\label{fig:workflow}
\end{figure*}

\section{Technique}

Given a target program, compiler flag selection decides which optimization flags to enable or disable so that the compiled program runs faster than the default \verb|-O3| setting. Many tuners provide useful search guidance, but this guidance is usually fixed once online measurement begins. \methodName{} instead represents the guidance as a trainable flag policy: offline pretraining initializes it from historical tuning data, and online tuning adapts it with feedback from the target program.

Figure~\ref{fig:workflow} presents an overview of \methodName{}, which comprises two stages: (1) offline pretraining and (2) online tuning. The offline stage consumes historical tuning records to train a single program-conditioned flag policy that can be reused across target programs. For each new target program, this policy recommends a program-specific starting point in the flag space for online tuning. The online stage starts from this point and updates the same policy on the target program through real compile-run feedback.

\subsection{Offline Pretraining}
\label{sec:offline-pretraining}

To give each target program its own starting point rather than a generic, program-agnostic one, \methodName{} uses the offline stage to turn historical tuning experience into a program-conditioned distribution over the flag space.
To this end, \methodName{} casts offline learning as supervised policy pretraining over the full flag space, rather than treating historical data as a one-shot performance predictor or a fixed search bias as in prior work. 
The offline input consists of historical tuning records, each containing a source program, a measured binary flag configuration, and the corresponding speedup over \verb|-O3|. These records are not tied to any single tuner: they can be collected from existing techniques such as RIO~\cite{rio}, PDCAT~\cite{pdcat}, or other search procedures. 
From these records, \methodName{} first represents each program from its source code and then uses good historical configurations as supervised signals to pretrain a flag policy with per-flag binary cross-entropy.

\paragraph{Program Representation.}

To produce a program-specific starting point, \methodName{} first converts each source program into an embedding.  This embedding is later passed to the flag policy so that flag probabilities depend on the target program.

\methodName{} builds this embedding from source code using CodeBERT. This design avoids relying on GCC-specific IR features or pass traces, and keeps feature extraction independent of compiler internals. Source code exposes structures that are relevant to optimization decisions, such as loops, branches, function calls, and memory accesses. A pretrained code model provides a compact way to encode these source-level signals.

Since a complete program may exceed the input length supported by CodeBERT, \methodName{} first concatenates the C source files of the program and splits the text into bounded-length chunks. Each chunk is tokenized and truncated to at most 512 tokens~\cite{codebert}. \methodName{} encodes each chunk with CodeBERT and uses the hidden state of the \verb|[CLS]| token as the chunk embedding.

To obtain a single embedding for the whole program, \methodName{} averages the chunk embeddings:
\begin{equation}
    c_p = \frac{1}{m}\sum_{j=1}^{m} h_j .
\end{equation}

Here, $h_j \in \mathbb{R}^{768}$ denotes the embedding of the $j$-th chunk, and $m$ is the number of chunks for program $p$. Averaging chunk embeddings gives \methodName{} a program embedding $c_p$ regardless of program size. This embedding is used as the program input to the flag policy described next.

\paragraph{Supervised Pretraining.}

Given the program embedding $c_p$, \methodName{} trains a program-conditioned flag policy to estimate the enable probability of each optimization flag for program $p$. The policy also takes the \verb|-O3| flag state $x_0$ as input, since \methodName{} searches for configurations that improve over the compiler's default \verb|-O3| setting. The policy maps the program representation and the \verb|-O3| reference state to per-flag logits.

In our implementation, the same policy architecture is used in both offline pretraining and online tuning. Let $n$ be the number of optimization flags considered by \methodName{}. The \verb|-O3| flag vector is first encoded by a learnable embedding layer. For each optimization flag, the model maintains two 16-dimensional embeddings, corresponding to the disabled and enabled states, respectively. For each flag, the embedding corresponding to its \verb|-O3| state is selected, and the resulting flag-state embeddings are stacked and flattened. A linear layer with ReLU activation then maps the flattened vector to a 128-dimensional state representation. This state representation is concatenated with the program embedding and passed through a multilayer perceptron with three linear layers of dimensions $896 \rightarrow 512 \rightarrow 256 \rightarrow 16n$, where LayerNorm and ReLU are applied after the first two linear layers. The output is reshaped into an $n \times 16$ matrix, so that each optimization flag is associated with one 16-dimensional prediction vector. Finally, the logit $z_i$ of each flag is computed as the dot product between its 16-dimensional prediction vector and a trainable \verb|-O3| anchor embedding.
The enable probability of flag $i \in \{1,\ldots,n\}$ is then computed as
\begin{equation}
    q_i = \pi_\theta(x_i = 1 \mid c_p, x_0) = \sigma(z_i),
\end{equation}

where $\theta$ denotes the policy parameters, $x_i \in \{0,1\}$ is the binary decision for the $i$-th flag, and $\sigma(\cdot)$ is the sigmoid function. Here, $x_i=1$ means enabling the flag and $x_i=0$ means disabling it.

To construct supervision, \methodName{} selects historical configurations that improve over \verb|-O3| and treats their binary flag vectors as labels. Given a training pair $(p,y)$, where $y=(y_1,\ldots,y_n)$ is such a label vector for program $p$, the policy is trained with per-flag binary cross-entropy:
\begin{equation}
    \mathcal{L}_{\mathrm{pre}}(\theta)
    =
    -\sum_{i=1}^{n}
    \left[
    y_i \log q_i + (1-y_i)\log(1-q_i)
    \right].
\end{equation}
Here, $y_i$ is the historical label for the $i$-th flag.

The resulting pretrained policy provides the program-specific starting point for online tuning.

\subsection{Online Tuning}
\label{sec:online-grpo-tuning}

Because the effects of optimization flags are program-dependent, the policy learned from historical data may not fully match a new target program. \methodName{} therefore refines the policy online using real compile-run feedback. We instantiate this online update with GRPO because each tuning round evaluates a group of candidate configurations for the same program, making their relative performance directly comparable.

Since compiler flag tuning decides whether each optimization flag should be enabled or disabled, \methodName{} models a complete configuration as a binary decision vector. Let $x=(x_1,\ldots,x_n)$ denote a candidate flag configuration, where $n$ is the number of optimization flags and $x_i \in \{0,1\}$ indicates whether the $i$-th flag is enabled. Given the target program embedding $c_p$ and the \verb|-O3| flag state $x_0$, the policy parameterized by $\theta$ outputs an enable probability $q_i$ for each flag. These probabilities define a factorized Bernoulli policy:
\begin{equation}
    \pi_\theta(x \mid c_p,x_0) =
    \prod_{i=1}^{n} q_i^{x_i}(1-q_i)^{1-x_i},
\end{equation}
where $\pi_\theta(x \mid c_p,x_0)$ is the probability of sampling the complete configuration $x$ under the current policy.

In each online round, \methodName{} samples $K$ configurations $\{x^{(1)},\ldots,x^{(K)}\}$ from the current policy. Each sampled binary vector is translated into compiler options: bit 1 enables the corresponding \verb|-f...| flag, while bit 0 uses the \verb|-fno-...| form.

\methodName{} then compiles and executes the target program with each candidate configuration. The reward is defined as speedup over \verb|-O3|, with failed measurements penalized. For candidate $x^{(k)}$, let $T^{(k)}$ be its runtime, $T_{O3}$ be the runtime under \verb|-O3|, and $m^{(k)}$ indicate whether compilation and execution succeed. The reward is
\begin{equation}
    r^{(k)} =
    \begin{cases}
    T_{O3}/T^{(k)}, & m^{(k)}=1,\\
    -1, & m^{(k)}=0.
    \end{cases}
\end{equation}
Thus, $r^{(k)}>1$ means the candidate is faster than \verb|-O3|, while failed candidates receive reward $-1$.

Since all $K$ candidates in a round are evaluated on the same program with the same metric, their rewards can be compared directly. \methodName{} uses this group-relative comparison to compute the advantage without training a separate value model:
\begin{equation}
    A^{(k)} =
    \frac{r^{(k)} - \mathrm{mean}(r^{(1)},\ldots,r^{(K)})}
    {\mathrm{std}(r^{(1)},\ldots,r^{(K)})+\delta},
\end{equation}
where $A^{(k)}$ is the advantage of candidate $x^{(k)}$, and $\delta$ is a small constant for numerical stability. A positive value means the candidate performs better than the group average, while a negative value means the opposite.

The policy should move toward candidates with positive advantages, but each update is based on only one measured group and should therefore be constrained. We use a clipped GRPO/PPO-style objective~\cite{deepseekmath_grpo,ppo}. Let $\theta_{\mathrm{old}}$ be the policy parameters used to sample the current group. For candidate $x^{(k)}$, the probability ratio between the current policy and the sampling policy is
\begin{equation}
    \rho^{(k)}(\theta)
    =
    \frac{
    \pi_\theta(x^{(k)} \mid c_p,x_0)
    }{
    \pi_{\theta_{\mathrm{old}}}(x^{(k)} \mid c_p,x_0)
    }.
\end{equation}

To keep the objective readable, we first define the clipped policy-improvement term for each candidate:
\begin{equation}
\begin{aligned}
    g^{(k)}(\theta) =
    \min\big(&
    \rho^{(k)}(\theta)A^{(k)},\\
    &
    \mathrm{clip}(\rho^{(k)}(\theta),1-\epsilon,1+\epsilon)A^{(k)}
    \big).
\end{aligned}
\end{equation}
Here, $\epsilon$ is the clipping threshold.

The online loss aggregates this clipped signal over the sampled group and adds entropy regularization:
\begin{equation}
    \mathcal{L}_{\mathrm{online}}(\theta)
    =
    -\frac{1}{K}\sum_{k=1}^{K} g^{(k)}(\theta)
    - \beta H(\pi_\theta).
\end{equation}
Here, $H(\pi_\theta)$ is the entropy of the Bernoulli policy, and $\beta$ controls the entropy bonus. Minimizing this loss increases the probability of better-than-average configurations while clipping limits overly large policy changes.

Online tuning repeats this process until the tuning budget is exhausted. \methodName{} returns the best measured configuration $x_{\mathrm{best}}$ and its speedup $r_{\mathrm{best}}$.

\section{Experimental Setup}

To evaluate the performance of \methodName{}, this experiment is designed to answer two research questions.

\begin{itemize}
    \item \textbf{RQ1: Effectiveness.} How does \methodName{} perform compared to representative compiler auto-tuning techniques? This RQ investigates the performance of flag configurations selected by different compiler auto-tuning techniques.

    \item \textbf{RQ2: Ablation analysis.} How do the main components and design choices of \methodName{} contribute to the performance? This RQ investigates the impact of offline pretraining, program representation, the choice of online policy optimizer, GRPO group size, and different historical tuning data sources.
\end{itemize}

\subsection{Dataset}

Following existing works~\cite{pdcat,srtuner,grouptuner,cfsca}, we use GCC 15.2.0, the latest GCC release available at the time the experiments were conducted, as the target compiler, and evaluate \methodName{} on two widely used C benchmarks, cBench~\cite{cbench} and PolyBench~\cite{polybench}, which cover a wide range of practical functions ranging from 100 to 26,000 lines of code. For compiler flag tuning, we consider 126 optimization flags supported by GCC. Table~\ref{tab:benchmarks} summarizes the target programs used in our experiments, where the ``ID'' column uses C1--C20 to denote the 20 cBench programs and P1--P30 to denote the 30 PolyBench programs.

\begin{table}[t]
\caption{Statistics of Programs from cBench and PolyBench.}
\label{tab:benchmarks}
\centering
\footnotesize
\setlength{\tabcolsep}{4pt}
\begin{tabular}{@{}c l r @{\hspace{1.6em}} c l r@{}}
\toprule
\textbf{ID} & \textbf{Program} & \textbf{Lines} &
\textbf{ID} & \textbf{Program} & \textbf{Lines} \\
\midrule
C1 & automotive\_bitcount & 954 & C11 & consumer\_jpeg\_d & 26,183 \\
C2 & automotive\_susan\_e & 2,129 & C12 & consumer\_tiff2bw & 22,271 \\
C3 & automotive\_susan\_c & 2,129 & C13 & consumer\_tiffdither & 22,184 \\
C4 & automotive\_susan\_s & 2,129 & C14 & consumer\_tiffmedian & 22,756 \\
C5 & consumer\_tiff2rgba & 22,321 & C15 & network\_dijkstra & 199 \\
C6 & consumer\_jpeg\_c & 26,950 & C16 & network\_patricia & 634 \\
C7 & security\_sha & 297 & C17 & security\_blowfish\_d & 1,524 \\
C8 & bzip2e & 7,200 & C18 & security\_blowfish\_e & 1,524 \\
C9 & telecom\_adpcm\_c & 389 & C19 & telecom\_CRC32 & 307 \\
C10 & bzip2d & 7,200 & C20 & automotive\_qsort1 & 227 \\
\midrule
P1 & correlation & 248 & P16 & syrk & 199 \\
P2 & covariance & 218 & P17 & trmm & 199 \\
P3 & symm & 231 & P18 & atax & 198 \\
P4 & 2mm & 252 & P19 & bicg & 214 \\
P5 & 3mm & 267 & P20 & doitgen & 203 \\
P6 & cholesky & 212 & P21 & mvt & 210 \\
P7 & lu & 210 & P22 & durbin & 195 \\
P8 & nussinov & 569 & P23 & gramschmidt & 220 \\
P9 & heat-3d & 211 & P24 & ludcmp & 247 \\
P10 & jacobi-2d & 200 & P25 & trisolv & 183 \\
P11 & gemm & 221 & P26 & deriche & 265 \\
P12 & gemver & 249 & P27 & floyd-warshall & 176 \\
P13 & gesummv & 210 & P28 & adi & 237 \\
P14 & seidel-2d & 179 & P29 & fdtd-2d & 245 \\
P15 & syr2k & 214 & P30 & jacobi-1d & 186 \\
\bottomrule
\end{tabular}
\end{table}

\subsection{Compared Techniques}

To validate the effectiveness of our technique, we compare \methodName{} against four representative compiler auto-tuning techniques: (1) \textbf{RIO}~\cite{rio}, which is a random iterative optimization technique and is commonly used as a baseline in compiler auto-tuning studies, (2) \textbf{SRTuner}~\cite{srtuner}, which is a search-based auto-tuning technique that exploits synergistic relations among optimization flags to guide the search process, (3) \textbf{GroupTuner}~\cite{grouptuner}, which is a group-aware technique that organizes flags into coherent option groups and performs localized mutation around historically good configurations, and (4) \textbf{PDCAT}~\cite{pdcat}, which is a preference-driven compiler auto-tuning technique that reduces the search space using constraints and preferences. In particular, GroupTuner and PDCAT are among the most recently published techniques and represent the current state of the art in flag selection.

For a fair comparison, we use a time budget as the termination condition of the compared techniques and compare the flag configurations generated by these techniques. We report results at 1,250, 2,500, and 5,000 seconds, respectively, to compare their performance. Moreover, we repeat each technique five times to alleviate the influence of randomness
resulting from the compiler auto-tuning techniques and use the median result to denote the performance of an auto-tuning technique.
\begin{table*}[htbp]
\caption{Speedup over \texttt{-O3} achieved by compared techniques.}
\label{tab:overall-effectiveness-full}
\centering
\scriptsize
\setlength{\tabcolsep}{1pt}
\renewcommand{\arraystretch}{0.95}
\resizebox{\textwidth}{!}{
\begin{tabular}{l||c|ccc|c|ccc|c|ccc|c|ccc|c|ccc}
\toprule
\textbf{Technique}
& \textbf{ID} & \textbf{1,250s} & \textbf{2,500s} & \textbf{5,000s}
& \textbf{ID} & \textbf{1,250s} & \textbf{2,500s} & \textbf{5,000s}
& \textbf{ID} & \textbf{1,250s} & \textbf{2,500s} & \textbf{5,000s}
& \textbf{ID} & \textbf{1,250s} & \textbf{2,500s} & \textbf{5,000s}
& \textbf{ID} & \textbf{1,250s} & \textbf{2,500s} & \textbf{5,000s} \\
\midrule
\methodName{} & \multirow{5}{*}{C1 (3)} & 1.700 & 1.734 & 1.759 & \multirow{5}{*}{C2 (1)} & 1.525 & 1.541 & 1.585 & \multirow{5}{*}{C3 (1)} & 1.400 & 1.425 & 1.458 & \multirow{5}{*}{C5 (1)} & 1.251 & 1.336 & 1.392 & \multirow{5}{*}{C6 (3)} & 1.334 & 1.438 & 1.518 \\
RIO & & \underline{1.512} & \underline{1.571} & \underline{1.606} & & \underline{1.417} & \underline{1.465} & \underline{1.545} & & 1.347 & 1.399 & 1.485 & & 1.420 & 1.420 & 1.420 & & 1.108 & 1.207 & 1.323 \\
SRTuner & & 1.796 & 1.839 & 1.863 & & 1.581 & 1.581 & 1.662 & & 1.414 & 1.414 & 1.414 & & \underline{1.178} & \underline{1.222} & \underline{1.222} & & \underline{1.074} & \underline{1.137} & \underline{1.214} \\
GroupTuner & & \textbf{1.856} & \textbf{1.896} & \textbf{1.962} & & 1.525 & 1.721 & 1.723 & & \underline{1.300} & \underline{1.377} & \underline{1.402} & & \textbf{1.442} & \textbf{1.541} & \textbf{1.618} & & 1.401 & \textbf{1.576} & \textbf{1.624} \\
PDCAT & & 1.853 & 1.884 & 1.896 & & \textbf{1.671} & \textbf{1.799} & \textbf{1.799} & & \textbf{1.559} & \textbf{1.564} & \textbf{1.564} & & 1.313 & 1.503 & 1.575 & & \textbf{1.470} & 1.515 & 1.570 \\
\midrule
\methodName{} & \multirow{5}{*}{C7 (1)} & 2.059 & 2.153 & \underline{2.153} & \multirow{5}{*}{C8 (3)} & \textbf{1.500} & \textbf{1.516} & \textbf{1.564} & \multirow{5}{*}{C9 (1)} & \textbf{3.012} & \textbf{3.090} & 3.162 & \multirow{5}{*}{C12 (1)} & \textbf{1.458} & \textbf{1.458} & 1.494 & \multirow{5}{*}{C14 (1)} & 1.444 & 1.460 & \textbf{1.623} \\
RIO & & 2.116 & 2.279 & 2.376 & & 1.270 & 1.382 & 1.442 & & 2.661 & 2.661 & 2.883 & & 1.254 & 1.323 & 1.323 & & 1.386 & 1.386 & 1.461 \\
SRTuner & & \underline{2.017} & \underline{2.035} & 2.320 & & \underline{1.241} & \underline{1.284} & \underline{1.299} & & 2.565 & 2.623 & 2.623 & & \underline{1.167} & \underline{1.167} & \underline{1.167} & & \underline{1.297} & \underline{1.307} & \underline{1.307} \\
GroupTuner & & 2.178 & \textbf{2.489} & \textbf{2.638} & & 1.425 & 1.435 & 1.487 & & 2.538 & 2.894 & \textbf{3.375} & & 1.400 & 1.427 & \textbf{1.618} & & 1.371 & 1.371 & 1.464 \\
PDCAT & & \textbf{2.319} & 2.319 & 2.481 & & 1.384 & 1.400 & 1.518 & & \underline{1.578} & \underline{1.602} & \underline{1.699} & & 1.324 & 1.427 & 1.599 & & \textbf{1.486} & \textbf{1.510} & 1.510 \\
\midrule
\methodName{} & \multirow{5}{*}{C15 (1)} & 1.553 & 1.586 & \textbf{1.667} & \multirow{5}{*}{C18 (2)} & \textbf{1.883} & \textbf{1.974} & \textbf{2.047} & \multirow{5}{*}{C20 (1)} & 1.692 & \textbf{1.723} & \textbf{1.888} & \multirow{5}{*}{P1 (1)} & \textbf{1.224} & 1.251 & 1.284 & \multirow{5}{*}{P2 (2)} & \textbf{1.224} & 1.235 & 1.249 \\
RIO & & \underline{1.315} & 1.440 & 1.528 & & 1.651 & 1.672 & 1.732 & & 1.600 & 1.666 & 1.671 & & 1.196 & 1.221 & 1.243 & & 1.209 & 1.210 & 1.226 \\
SRTuner & & 1.439 & \underline{1.439} & \underline{1.489} & & 1.494 & \underline{1.494} & \underline{1.500} & & \underline{1.530} & \underline{1.530} & \underline{1.587} & & 1.174 & 1.203 & 1.203 & & 1.139 & 1.220 & 1.220 \\
GroupTuner & & 1.483 & 1.498 & 1.618 & & \underline{1.492} & 1.606 & 1.760 & & \textbf{1.706} & 1.713 & 1.719 & & \underline{1.044} & \underline{1.064} & \underline{1.165} & & \underline{1.055} & \underline{1.058} & \underline{1.110} \\
PDCAT & & \textbf{1.598} & \textbf{1.614} & 1.621 & & 1.655 & 1.684 & 1.859 & & 1.690 & 1.706 & 1.706 & & 1.190 & \textbf{1.317} & \textbf{1.317} & & 1.206 & \textbf{1.248} & \textbf{1.290} \\
\midrule
\methodName{} & \multirow{5}{*}{P4 (1)} & \textbf{1.525} & \textbf{1.532} & \textbf{1.570} & \multirow{5}{*}{P5 (1)} & \textbf{1.272} & \textbf{1.272} & \textbf{1.289} & \multirow{5}{*}{P11 (1)} & \textbf{1.866} & \textbf{1.941} & \textbf{2.233} & \multirow{5}{*}{P12 (1)} & 1.391 & 1.391 & 1.509 & \multirow{5}{*}{P15 (1)} & \textbf{1.398} & \textbf{1.402} & \textbf{1.402} \\
RIO & & 1.128 & \underline{1.128} & 1.165 & & 1.121 & 1.147 & 1.147 & & 1.202 & \underline{1.202} & 1.229 & & \underline{1.336} & \underline{1.336} & \underline{1.373} & & 1.221 & 1.227 & 1.278 \\
SRTuner & & \underline{1.090} & 1.133 & \underline{1.133} & & 1.141 & 1.141 & 1.175 & & \underline{1.195} & 1.204 & \underline{1.207} & & \textbf{1.462} & \textbf{1.462} & 1.470 & & 1.225 & 1.249 & 1.249 \\
GroupTuner & & 1.151 & 1.189 & 1.241 & & \underline{1.107} & 1.142 & 1.142 & & 1.285 & 1.298 & 1.352 & & 1.418 & 1.461 & \textbf{1.531} & & \underline{1.137} & \underline{1.137} & 1.255 \\
PDCAT & & 1.116 & 1.197 & 1.244 & & 1.108 & \underline{1.108} & \underline{1.124} & & 1.365 & 1.365 & 1.365 & & 1.342 & 1.342 & 1.394 & & 1.169 & 1.184 & \underline{1.194} \\
\midrule
\methodName{} & \multirow{5}{*}{P16 (1)} & \textbf{1.483} & \textbf{1.507} & \textbf{1.512} & \multirow{5}{*}{P17 (2)} & \textbf{1.284} & \textbf{1.296} & \textbf{1.313} & \multirow{5}{*}{P20 (2)} & 1.578 & 1.657 & 1.675 & \multirow{5}{*}{P21 (1)} & \textbf{1.668} & \textbf{1.700} & \textbf{1.828} & \multirow{5}{*}{P23 (3)} & \textbf{1.737} & \textbf{1.780} & 1.835 \\
RIO & & 1.237 & 1.237 & 1.304 & & 1.220 & 1.254 & 1.290 & & \underline{1.558} & \underline{1.567} & \underline{1.583} & & \underline{1.384} & \underline{1.384} & \underline{1.454} & & \underline{1.592} & \underline{1.652} & 1.755 \\
SRTuner & & \underline{1.190} & \underline{1.202} & \underline{1.202} & & 1.228 & 1.248 & 1.248 & & 1.670 & 1.680 & 1.682 & & 1.591 & 1.649 & 1.649 & & 1.621 & 1.670 & \underline{1.676} \\
GroupTuner & & 1.206 & 1.228 & 1.376 & & \underline{1.169} & \underline{1.227} & \underline{1.239} & & 1.651 & 1.706 & \textbf{1.767} & & 1.567 & 1.595 & 1.811 & & 1.610 & 1.697 & \textbf{1.860} \\
PDCAT & & 1.192 & 1.210 & 1.370 & & 1.208 & 1.242 & 1.275 & & \textbf{1.716} & \textbf{1.717} & 1.747 & & 1.478 & 1.478 & 1.478 & & 1.664 & 1.693 & 1.754 \\
\midrule
\methodName{} & \multirow{5}{*}{P24 (1)} & 1.368 & 1.383 & \textbf{1.416} & \multirow{5}{*}{P26 (1)} & 1.812 & 2.054 & 2.054 & \multirow{5}{*}{P27 (3)} & 1.098 & 1.116 & 1.133 & \multirow{5}{*}{P29 (1)} & 1.148 & 1.163 & 1.192 & \multirow{5}{*}{P30 (2)} & \textbf{3.583} & \textbf{3.712} & \textbf{4.159} \\
RIO & & 1.273 & 1.299 & 1.299 & & 1.473 & 1.566 & 1.584 & & \textbf{1.099} & 1.133 & 1.155 & & \underline{1.068} & \underline{1.068} & 1.131 & & 2.867 & 2.932 & 3.107 \\
SRTuner & & 1.200 & \underline{1.200} & \underline{1.200} & & \textbf{2.224} & \textbf{2.224} & \textbf{2.233} & & \underline{1.070} & \underline{1.087} & \underline{1.111} & & 1.102 & 1.126 & \underline{1.126} & & \underline{1.777} & \underline{1.895} & \underline{1.971} \\
GroupTuner & & \textbf{1.390} & \textbf{1.390} & 1.392 & & 1.398 & 1.434 & 1.644 & & 1.094 & 1.113 & \textbf{1.161} & & \textbf{1.154} & \textbf{1.238} & \textbf{1.258} & & 3.394 & 3.628 & 3.763 \\
PDCAT & & \underline{1.178} & 1.238 & 1.288 & & \underline{1.312} & \underline{1.317} & \underline{1.436} & & 1.095 & \textbf{1.140} & 1.142 & & 1.083 & 1.111 & 1.166 & & 3.022 & 3.234 & 3.410 \\
\bottomrule
\end{tabular}
}
\end{table*}
\subsection{Implementation} 

We implement \methodName{} in Python based on PyTorch and Hugging Face Transformers~\cite{pytorch,transformers}. 

For offline pretraining, the learning rate is set to $5\times10^{-4}$, the batch size is set to 64, and the model is trained for 300 epochs. We use AdamW with weight decay $10^{-4}$ and CosineAnnealingLR with $\eta_{\min}=10^{-5}$. Gradient clipping is applied with $\texttt{max\_norm}=1.0$ during offline pretraining. In particular, the offline pretraining stage uses historical tuning data collected by random iterative optimization. After filtering configurations whose speedup is not larger than 1, the three training splits contain 29,004, 25,525, and 21,553 records, respectively.

In the GRPO instantiation of online tuning, the group size $K$ is set to 8, the online learning rate is set to $10^{-4}$, the entropy coefficient $\beta$ is set to 0.02, and the clipping threshold $\epsilon$ is set to 0.2. The numerical stability constant $\delta$ in advantage normalization is set to $10^{-8}$. Each online round performs one clipped policy update and applies gradient clipping with $\texttt{max\_norm}=0.5$.

Following the protocol in previous work~\cite{pdcat}, we perform three independent random splits of the programs. In each split, 70\% of the total programs (i.e., 35 programs) are used for offline pretraining and the remaining 30\% serve as testing programs for evaluation. We train one model for each split and evaluate it on the corresponding testing programs to reduce the influence of a particular train-test selection.

For the compared techniques, we directly use their official reproducible package when available (i.e., for SRTuner, GroupTuner and PDCAT); otherwise, we re-implement the technique (i.e., RIO) strictly following the description in the paper. The experiments are performed on a server with two Intel Xeon Platinum 8481C CPUs, 1.0 TiB memory, one NVIDIA GeForce RTX 4090 D GPU, and Ubuntu 22.04.3 LTS.

\subsection{Measurement}

Following prior work~\cite{pdcat,srtuner,grouptuner,cfsca}, we report speedup over GCC \verb|-O3|. GCC also provides \verb|-O1|, \verb|-O2|, and \verb|-Ofast|, but \verb|-O3| is the highest standard optimization level and is commonly used as the baseline in compiler tuning studies.

To obtain the runtime performance improvement provided by a compiler auto-tuning technique, we compile any given program $P$ using both the selected optimization sequence from the auto-tuning technique and the default \verb|-O3| optimization. This process yields two compiled programs $P_s$ and $P_d$. Then we execute both programs and measure the speedup of $P$ by dividing the execution time of $P_d$ (i.e., compiled with \verb|-O3|) by the execution time of $P_s$ (i.e., compiled with the selected optimization sequence). A speedup larger than 1 indicates that the selected optimization sequence outperforms \verb|-O3|. We use this result to measure compiler auto-tuning techniques.

To reduce measurement noise, \methodName{} measures each candidate configuration and the \verb|-O3| baseline in the same worker and in adjacent time windows. Each worker uses an independent temporary directory that isolates intermediate files during the parallel evaluation of candidates.

\section{Experimental Results}
\label{sec:experimental-results}
\subsection{Overall Effectiveness (RQ1)}
\label{sec:overall}

To answer RQ1, we compare \methodName{} with representative compiler auto-tuning techniques under the same tuning budgets, using speedup over GCC \verb|-O3| as the metric. Table~\ref{tab:overall-effectiveness-full} presents the program-wise speedups achieved by the compared techniques. Each cell reports the speedup over \verb|-O3|, where a larger value indicates better runtime performance. For each target program, the best result is highlighted in bold and the worst is underlined. When a program appears in multiple train-test selections, the table reports its average speedup and the number of appearances.

As shown in the table, \methodName{} improves over GCC \verb|-O3| on all target programs, with the speedup ranging from $1.133\times$ to $4.159\times$ in 5,000 seconds. Its performance also improves as the tuning budget increases, suggesting that online tuning continues to refine the configurations sampled from the pretrained policy.

\subsubsection{Statistics Analysis}

To further examine whether \methodName{} outperforms the compared baselines, we statistically analyze the results from five perspectives. First, we count how often each baseline performs worse than \methodName{} under different tuning budgets. Second, we count the number of best and worst results achieved by each technique, which reflects both effectiveness and robustness. Third, we compare the average speedup across all testing programs to measure the overall acceleration performance. Fourth, we conduct paired statistical tests between \methodName{} and each baseline. Finally, we report the offline pretraining cost introduced by \methodName{}.

Comparing \methodName{} against each baseline program by program, we find that most baselines produce less effective flag configurations under the same tuning budget. Specifically, RIO, SRTuner, GroupTuner, and PDCAT perform worse than \methodName{} on 27/27/26, 24/25/25, 20/21/18, and 21/18/19 of the 30 programs in 1,250/2,500/5,000 seconds, respectively.

\begin{table}[htbp]
\caption{Number of best and worst performance results.}
\label{tab:best-worst-summary}
\centering
\small
\setlength{\tabcolsep}{6pt}
\renewcommand{\arraystretch}{1.0}
\begin{tabular}{lccc}
\toprule
\multicolumn{4}{c}{\textbf{Best results/Worst results}} \\
\midrule
\textbf{Technique} & \textbf{1,250s} & \textbf{2,500s} & \textbf{5,000s} \\
\midrule
\methodName{} & \textbf{15/0} & \textbf{14/0} & \textbf{14/1} \\
RIO        & 1/8  & 0/9  & 0/5  \\
SRTuner    & 2/12 & 2/13 & 1/16 \\
GroupTuner & 5/7  & 6/5  & 11/4 \\
PDCAT      & 7/3  & 8/3  & 4/4  \\
\bottomrule
\end{tabular}
\end{table}

We further count, for each case, how often each technique achieves the best result (in \textbf{bold}) and the worst result (\underline{underlined}). As summarized in Table~\ref{tab:best-worst-summary}, \methodName{} obtains the largest number of best results at all three budgets (15/14/14 under 1,250s/2,500s/5,000s) and rarely the worst (only 0/0/1). In contrast, RIO and SRTuner produce more worse results (8/9/5 and 12/13/16, respectively). GroupTuner and PDCAT are more competitive, but still give more worst cases than \methodName{} (7/5/4 and 3/3/4). These counts show that \methodName{} performs consistently across programs, rather than excelling only in a small subset.

In terms of average speedup, \methodName{} also ranks first under all three budgets, reaching $1.616\times$, $1.661\times$, and $1.732\times$, respectively. 
Among the strongest baselines, GroupTuner reaches $1.498\times$, $1.572\times$, and $1.669\times$, while PDCAT reaches $1.478\times$, $1.522\times$, and $1.580\times$ under the same budgets. The other two baselines also remain lower; even at the largest budget, RIO and SRTuner reach only $1.537\times$ and $1.481\times$, respectively. These results show that \methodName{} consistently improves average tuning quality, with its advantage remaining visible even against the strongest competing tuners.

 We further conduct a paired $t$-test on the 30 testing programs in 5,000 seconds. \methodName{} significantly outperforms RIO, SRTuner, and PDCAT, with p-values of $3.98\times10^{-4}$, $3.87\times10^{-3}$, and $2.88\times10^{-2}$, respectively. The improvement over GroupTuner, the strongest baseline, is not statistically significant ($p=1.74\times10^{-1}$); nevertheless, \methodName{} still achieves a higher average speedup and more best (and fewer worst) results.

Since \methodName{} introduces an offline pretraining stage that the other techniques do not have, we therefore report the offline pretraining cost of \methodName{} to discuss whether this stage incurs a large additional cost. Across the three random splits, supervised pretraining takes 9.40, 9.70, and 8.44 minutes, with an average of 9.18 minutes. Moreover, this cost is paid only once per split, after which the trained model is reused for all testing programs in that split, so it is negligible compared with the online tuning budget.

To sum up, \methodName{} improves over GCC \verb|-O3| on all testing programs and achieves the best overall performance among the compared techniques in terms of average speedup and best-result counts. This demonstrates the effectiveness of combining offline supervised pretraining with online reinforcement-learning-based policy optimization for compiler flag auto-tuning.

\subsubsection{Case Study}

To examine whether the best configuration selected by \methodName{} is consistent with the target program's characteristics, we inspect its tuning result on \texttt{jacobi-1d}. As shown in Table~\ref{tab:overall-effectiveness-full}, \methodName{} reaches a summarized speedup of $4.159\times$ on this program in 5,000 seconds. \texttt{jacobi-1d} is a stencil kernel dominated by repeated one-dimensional array updates, so loop and floating-point optimizations are likely to affect its runtime.

In the inspected tuning log, the best measured configuration achieves a speedup of $4.940\times$. Among its enabled flags, we focus on \verb|-ftree-loop-vectorize|, \verb|-funroll-loops|, and \verb|-ffast-math|, because their roles in GCC correspond to loop vectorization, loop unrolling, and aggressive floating-point optimization, respectively~\cite{gcc15_optimize_options}.

\begin{table}[htbp]
\caption{Influence of Selected Flags in the Tuning Log of \texttt{jacobi-1d}.}
\label{tab:case-study-flags}
\centering
\scriptsize
\setlength{\tabcolsep}{6pt}
\renewcommand{\arraystretch}{1.08}
\begin{tabular}{@{}lrr@{}}
\toprule
\textbf{Flag} & \textbf{Avg. On} & \textbf{Avg. Off} \\
\midrule
\texttt{-ftree-loop-vectorize} & 1.030 & 0.727 \\
\texttt{-funroll-loops}        & 1.027 & 0.852 \\
\texttt{-ffast-math}           & 1.026 & 0.861 \\
\bottomrule
\end{tabular}
\end{table}

As shown in Table~\ref{tab:case-study-flags}, candidates enabling these three flags consistently achieve higher average speedups than those disabling them. The clearest gap appears for \verb|-ftree-loop-vectorize|, whose average speedup is $1.030\times$ when enabled but only $0.727\times$ when disabled. Similar trends appear for the other two flags. Although some candidates without these flags still outperform \verb|-O3|, the enabled group performs better on average, suggesting that their appearance in the best configuration is not accidental.

These observations are consistent with the structure of \texttt{jacobi-1d}, whose inner loops repeatedly update contiguous floating-point arrays. This case study therefore suggests that \methodName{} can select optimization flags whose roles match the computation pattern of the target program.

\begin{table}[htbp]
\caption{Impact of offline pretraining.}
\label{tab:ablation-no-pretrain}
\centering
\scriptsize
\setlength{\tabcolsep}{3pt}
\renewcommand{\arraystretch}{1.05}
\begin{tabular}{l||c|ccc|c|ccc}
\toprule
\textbf{Technique}
& \textbf{ID} & \textbf{1,250s} & \textbf{2,500s} & \textbf{5,000s}
& \textbf{ID} & \textbf{1,250s} & \textbf{2,500s} & \textbf{5,000s} \\
\midrule
\methodName{} & \multirow{2}{*}{C2} & \textbf{1.525} & \textbf{1.541} & \textbf{1.585} & \multirow{2}{*}{C3} & \textbf{1.400} & \textbf{1.425} & \textbf{1.458} \\
\methodNoPre & & 1.440 & 1.500 & 1.578 & & 1.335 & 1.385 & 1.452 \\
\midrule
\methodName{} & \multirow{2}{*}{C5} & 1.251 & 1.336 & \textbf{1.392} & \multirow{2}{*}{C7} & \textbf{2.059} & \textbf{2.153} & \textbf{2.153} \\
\methodNoPre & & \textbf{1.313} & \textbf{1.359} & 1.383 & & 1.900 & 2.103 & 2.151 \\
\midrule
\methodName{} & \multirow{2}{*}{C9} & \textbf{3.012} & \textbf{3.090} & \textbf{3.162} & \multirow{2}{*}{C12} & \textbf{1.458} & 1.458 & 1.494 \\
\methodNoPre & & 2.977 & 3.037 & 3.151 & & 1.366 & \textbf{1.480} & \textbf{1.571} \\
\midrule
\methodName{} & \multirow{2}{*}{C14} & \textbf{1.444} & \textbf{1.460} & 1.623 & \multirow{2}{*}{C15} & \textbf{1.553} & \textbf{1.586} & \textbf{1.667} \\
\methodNoPre & & 1.370 & 1.420 & \textbf{1.819} & & 1.450 & 1.508 & 1.526 \\
\midrule
\methodName{} & \multirow{2}{*}{C20} & 1.692 & \textbf{1.723} & \textbf{1.888} & \multirow{2}{*}{P1} & 1.224 & \textbf{1.251} & \textbf{1.284} \\
\methodNoPre & & \textbf{1.702} & 1.713 & 1.713 & & \textbf{1.240} & 1.246 & 1.250 \\
\midrule
\methodName{} & \multirow{2}{*}{P4} & \textbf{1.525} & \textbf{1.532} & 1.570 & \multirow{2}{*}{P5} & 1.272 & 1.272 & 1.289 \\
\methodNoPre & & 1.463 & 1.489 & \textbf{1.583} & & \textbf{1.282} & \textbf{1.282} & \textbf{1.304} \\
\midrule
\methodName{} & \multirow{2}{*}{P11} & \textbf{1.866} & \textbf{1.941} & \textbf{2.233} & \multirow{2}{*}{P12} & \textbf{1.391} & 1.391 & \textbf{1.509} \\
\methodNoPre & & 1.750 & 1.880 & 2.225 & & 1.386 & \textbf{1.478} & 1.478 \\
\midrule
\methodName{} & \multirow{2}{*}{P15} & \textbf{1.398} & \textbf{1.402} & 1.402 & \multirow{2}{*}{P16} & \textbf{1.483} & \textbf{1.507} & 1.512 \\
\methodNoPre & & 1.379 & 1.398 & \textbf{1.477} & & 1.438 & 1.500 & \textbf{1.573} \\
\midrule
\methodName{} & \multirow{2}{*}{P21} & \textbf{1.668} & \textbf{1.700} & \textbf{1.828} & \multirow{2}{*}{P24} & \textbf{1.368} & \textbf{1.383} & \textbf{1.416} \\
\methodNoPre & & 1.528 & 1.547 & 1.567 & & 1.301 & 1.341 & 1.414 \\
\midrule
\methodName{} & \multirow{2}{*}{P26} & \textbf{1.812} & \textbf{2.054} & \textbf{2.054} & \multirow{2}{*}{P29} & 1.148 & 1.163 & 1.192 \\
\methodNoPre & & 1.373 & 1.403 & 1.426 & & \textbf{1.171} & \textbf{1.387} & \textbf{1.411} \\
\bottomrule
\end{tabular}
\end{table}

\subsection{Ablation Analysis (RQ2)}
To answer RQ2, we construct 7 variants of \methodName{} by removing or replacing each component individually or modifying parameters. We denote each variant as \methodName{} with a subscript indicating the modified component. We analyze the contribution of each component of \methodName{} by comparing the acceleration performance between \methodName{} and these variants in 1,250, 2,500, and 5,000 seconds.\footnote{In RQ1, we conduct three separate random selections of the initial tuning set to control the influence of randomness but achieve a consistent conclusion across
different selections. In this ablation analysis, we use one selection result as initial tuning and testing programs to control the expensive evaluation costs resulting from the 7 variants. The other settings remain the same as in Section~\ref{sec:overall}.}

\subsubsection{Offline Pretraining}

To investigate the contribution of offline pretraining, we replace the supervised pretraining stage by starting online tuning from a randomly initialized policy, resulting in a variant denoted as \methodNoPre. Table~\ref{tab:ablation-no-pretrain} shows that \methodName{} performs better than \methodNoPre{} on 15/15/13 out of 20 programs in 1,250/2,500/5,000 seconds. The average speedup gains are +0.069, +0.046, and +0.033. This result indicates that offline pretraining provides a useful initial policy, enabling online tuning to sample from more promising flag distributions before enough target-program feedback has been collected.

\subsubsection{Program Representation}

\begin{table}[htbp]
\caption{Impact of Program Representation.}
\label{tab:ablation-no-feature}
\centering
\scriptsize
\setlength{\tabcolsep}{3pt}
\renewcommand{\arraystretch}{1.05}
\begin{tabular}{l||c|ccc|c|ccc}
\toprule
\textbf{Technique}
& \textbf{ID} & \textbf{1,250s} & \textbf{2,500s} & \textbf{5,000s}
& \textbf{ID} & \textbf{1,250s} & \textbf{2,500s} & \textbf{5,000s} \\
\midrule
\methodName{} & \multirow{2}{*}{C2} & 1.525 & 1.541 & 1.585 & \multirow{2}{*}{C3} & 1.400 & 1.425 & 1.458 \\
\methodNoRep & & \textbf{1.633} & \textbf{1.633} & \textbf{1.753} & & \textbf{1.474} & \textbf{1.518} & \textbf{1.553} \\
\midrule
\methodName{} & \multirow{2}{*}{C5} & 1.251 & 1.336 & 1.392 & \multirow{2}{*}{C7} & 2.059 & 2.153 & 2.153 \\
\methodNoRep & & \textbf{1.334} & \textbf{1.372} & \textbf{1.433} & & \textbf{2.178} & \textbf{2.178} & \textbf{2.186} \\
\midrule
\methodName{} & \multirow{2}{*}{C9} & \textbf{3.012} & \textbf{3.090} & \textbf{3.162} & \multirow{2}{*}{C12} & \textbf{1.458} & \textbf{1.458} & 1.494 \\
\methodNoRep & & 2.796 & 2.836 & 2.937 & & 1.435 & 1.454 & \textbf{1.500} \\
\midrule
\methodName{} & \multirow{2}{*}{C14} & \textbf{1.444} & 1.460 & \textbf{1.623} & \multirow{2}{*}{C15} & \textbf{1.553} & \textbf{1.586} & \textbf{1.667} \\
\methodNoRep & & 1.396 & \textbf{1.468} & 1.482 & & 1.529 & 1.529 & 1.582 \\
\midrule
\methodName{} & \multirow{2}{*}{C20} & \textbf{1.692} & \textbf{1.723} & \textbf{1.888} & \multirow{2}{*}{P1} & 1.224 & 1.251 & 1.284 \\
\methodNoRep & & 1.586 & 1.675 & 1.694 & & \textbf{1.269} & \textbf{1.269} & \textbf{1.290} \\
\midrule
\methodName{} & \multirow{2}{*}{P4} & \textbf{1.525} & \textbf{1.532} & \textbf{1.570} & \multirow{2}{*}{P5} & \textbf{1.272} & 1.272 & 1.289 \\
\methodNoRep & & 1.431 & 1.518 & 1.518 & & 1.241 & \textbf{1.277} & \textbf{1.320} \\
\midrule
\methodName{} & \multirow{2}{*}{P11} & \textbf{1.866} & \textbf{1.941} & \textbf{2.233} & \multirow{2}{*}{P12} & \textbf{1.391} & \textbf{1.391} & \textbf{1.509} \\
\methodNoRep & & 1.760 & 1.930 & 2.215 & & 1.377 & 1.385 & 1.388 \\
\midrule
\methodName{} & \multirow{2}{*}{P15} & \textbf{1.398} & \textbf{1.402} & \textbf{1.402} & \multirow{2}{*}{P16} & 1.483 & 1.507 & 1.512 \\
\methodNoRep & & 1.280 & 1.372 & 1.380 & & \textbf{1.507} & \textbf{1.520} & \textbf{1.542} \\
\midrule
\methodName{} & \multirow{2}{*}{P21} & \textbf{1.668} & \textbf{1.700} & \textbf{1.828} & \multirow{2}{*}{P24} & \textbf{1.368} & \textbf{1.383} & \textbf{1.416} \\
\methodNoRep & & 1.557 & 1.687 & 1.691 & & 1.276 & 1.353 & 1.397 \\
\midrule
\methodName{} & \multirow{2}{*}{P26} & \textbf{1.812} & \textbf{2.054} & \textbf{2.054} & \multirow{2}{*}{P29} & \textbf{1.148} & \textbf{1.163} & 1.192 \\
\methodNoRep & & 1.693 & 1.693 & 1.693 & & 1.145 & 1.160 & \textbf{1.225} \\
\bottomrule
\end{tabular}
\end{table}

To examine the effect of program representation, we mask the CodeBERT source-code representation by replacing it with an all-zero vector of the same dimension, resulting in a variant denoted as \methodNoRep. In this variant, online search starts from the same initial flag distribution before program-specific feedback is observed. As reported in Table~\ref{tab:ablation-no-feature}, \methodName{} wins on 14/12/11 out of 20 programs across the three budgets and keeps a higher average speedup. At 5,000s, for example, \methodName{} reaches 1.686 on average, compared with 1.639 for \methodNoRep. This gap indicates that source-code features help the policy assign different starting distributions to different programs, rather than using the same search bias for all targets.

\subsubsection{Choice of Policy Optimizer}

To investigate the contribution of the GRPO update, we replace it with two alternative online reinforcement learning algorithms while keeping the pretrained policy unchanged. The first variant, denoted as \methodReinf, uses REINFORCE~\cite{williams1992simple} to directly update the policy with measured speedups as rewards. The second variant, denoted as \methodPPO, uses PPO~\cite{ppo} to perform clipped policy updates with advantage estimates from a learned value model.
Table~\ref{tab:ablation-reinforce} presents the acceleration performance of \methodName{}, \methodReinf, and \methodPPO{} on the selected test programs, with the best result in each case highlighted in bold.
\begin{table}[htbp]
\caption{Impact of Online Policy Optimizer.}
\label{tab:ablation-reinforce}
\centering
\scriptsize
\setlength{\tabcolsep}{3pt}
\renewcommand{\arraystretch}{1.05}
\begin{tabular}{l||c|ccc|c|ccc}
\toprule
\textbf{Technique}
& \textbf{ID} & \textbf{1,250s} & \textbf{2,500s} & \textbf{5,000s}
& \textbf{ID} & \textbf{1,250s} & \textbf{2,500s} & \textbf{5,000s} \\
\midrule
\methodName{} & \multirow{3}{*}{C2} & 1.525 & 1.541 & 1.585 & \multirow{3}{*}{C3} & 1.400 & 1.425 & 1.458 \\
\methodReinf & & \textbf{1.582} & \textbf{1.599} & \textbf{1.746} & & \textbf{1.487} & \textbf{1.632} & \textbf{1.632} \\
\methodPPO & & \underline{1.520} & \underline{1.536} & \underline{1.580} & & \underline{1.395} & \underline{1.420} & \underline{1.453} \\
\midrule
\methodName{} & \multirow{3}{*}{C5} & \underline{1.251} & \underline{1.336} & \underline{1.392} & \multirow{3}{*}{C7} & 2.059 & 2.153 & \underline{2.153} \\
\methodReinf & & 1.256 & \textbf{1.444} & \textbf{1.468} & & \textbf{2.165} & \textbf{2.350} & \textbf{2.496} \\
\methodPPO & & \textbf{1.311} & 1.357 & 1.396 & & \underline{1.936} & \underline{2.064} & 2.165 \\
\midrule
\methodName{} & \multirow{3}{*}{C9} & \textbf{3.012} & \textbf{3.090} & \textbf{3.162} & \multirow{3}{*}{C12} & \textbf{1.458} & \textbf{1.458} & 1.494 \\
\methodReinf & & \underline{2.557} & \underline{2.557} & \underline{2.689} & & \underline{1.238} & \underline{1.400} & \underline{1.400} \\
\methodPPO & & 2.827 & 2.999 & 3.117 & & 1.447 & 1.454 & \textbf{1.502} \\
\midrule
\methodName{} & \multirow{3}{*}{C14} & 1.444 & 1.460 & \textbf{1.623} & \multirow{3}{*}{C15} & 1.553 & 1.586 & \textbf{1.667} \\
\methodReinf & & \underline{1.300} & \underline{1.308} & \underline{1.383} & & \underline{1.403} & \underline{1.428} & \underline{1.499} \\
\methodPPO & & \textbf{1.489} & \textbf{1.508} & 1.565 & & \textbf{1.571} & \textbf{1.612} & 1.635 \\
\midrule
\methodName{} & \multirow{3}{*}{C20} & \textbf{1.692} & \textbf{1.723} & \textbf{1.888} & \multirow{3}{*}{P1} & 1.224 & \textbf{1.251} & \textbf{1.284} \\
\methodReinf & & 1.677 & 1.683 & 1.683 & & \underline{1.207} & \underline{1.207} & \underline{1.230} \\
\methodPPO & & \underline{1.652} & \underline{1.652} & \underline{1.665} & & \textbf{1.244} & 1.244 & 1.257 \\
\midrule
\methodName{} & \multirow{3}{*}{P4} & \textbf{1.525} & \textbf{1.532} & \textbf{1.570} & \multirow{3}{*}{P5} & \textbf{1.272} & 1.272 & 1.289 \\
\methodReinf & & \underline{1.127} & \underline{1.127} & \underline{1.184} & & \underline{1.115} & \underline{1.151} & \underline{1.173} \\
\methodPPO & & 1.470 & 1.525 & 1.525 & & 1.252 & \textbf{1.291} & \textbf{1.330} \\
\midrule
\methodName{} & \multirow{3}{*}{P11} & 1.866 & 1.941 & 2.233 & \multirow{3}{*}{P12} & 1.391 & \underline{1.391} & 1.509 \\
\methodReinf & & \underline{1.174} & \underline{1.241} & \underline{1.354} & & \underline{1.357} & 1.404 & \underline{1.425} \\
\methodPPO & & \textbf{2.015} & \textbf{2.015} & \textbf{2.305} & & \textbf{1.524} & \textbf{1.524} & \textbf{1.569} \\
\midrule
\methodName{} & \multirow{3}{*}{P15} & \textbf{1.398} & \textbf{1.402} & \underline{1.402} & \multirow{3}{*}{P16} & 1.483 & 1.507 & 1.512 \\
\methodReinf & & \underline{1.227} & \underline{1.276} & 1.413 & & \underline{1.215} & \underline{1.278} & \underline{1.306} \\
\methodPPO & & 1.354 & 1.399 & \textbf{1.448} & & \textbf{1.486} & \textbf{1.546} & \textbf{1.566} \\
\midrule
\methodName{} & \multirow{3}{*}{P21} & \textbf{1.668} & \textbf{1.700} & \textbf{1.828} & \multirow{3}{*}{P24} & \textbf{1.368} & 1.383 & 1.416 \\
\methodReinf & & \underline{1.435} & \underline{1.435} & \underline{1.472} & & \underline{1.237} & \underline{1.237} & \underline{1.309} \\
\methodPPO & & 1.550 & 1.611 & 1.611 & & 1.361 & \textbf{1.461} & \textbf{1.461} \\
\midrule
\methodName{} & \multirow{3}{*}{P26} & 1.812 & \textbf{2.054} & \textbf{2.054} & \multirow{3}{*}{P29} & \textbf{1.148} & \textbf{1.163} & \textbf{1.192} \\
\methodReinf & & \underline{1.301} & \underline{1.345} & \underline{1.362} & & \underline{1.060} & \underline{1.096} & \underline{1.101} \\
\methodPPO & & \textbf{1.842} & 2.043 & 2.049 & & 1.143 & 1.158 & 1.187 \\
\bottomrule
\end{tabular}
\end{table}

Among the three online RL algorithms, GRPO offers the best overall trade-off in our setting. \methodName{} outperforms \methodReinf{} on 16/15/15 of 20 programs in 1,250/2,500/5,000 seconds, with average speedups higher by +0.171, +0.159, and +0.169—indicating that using raw measured rewards directly, as REINFORCE does, yields a weaker and noisier update signal for compiler auto-tuning. 
Against \methodPPO, \methodName{} wins on 12/12/11 of 20 programs: PPO is competitive but provides no clear advantage here and additionally requires a separate value model that GRPO avoids by relying on within-group comparisons among candidates evaluated on the same program.
Overall, the offline-to-online framework is compatible with different online RL algorithms, but GRPO remains the preferred choice, as it matches the group structure of compiler tuning.

\subsubsection{Generality Across Data Sources}

To test the generality of \methodName{} across historical tuning data sources, we pretrain the policy with records collected by different tuning techniques. The main method in this paper, \methodName{}, is pretrained with records collected by random iterative optimization (RIO), while we further construct a variant, denoted as \methodPDCAT, using records collected by PDCAT. \methodPDCAT{} outperforms PDCAT on 25/21/22 out of 30 programs in 1,250/2,500/5,000 seconds, with average speedup gains of +0.123, +0.114, and +0.122. Meanwhile, \methodName{} outperforms RIO on 27/27/26 out of 30 programs, with gains of +0.174, +0.180, and +0.195. These results suggest that the framework can reuse historical records from different tuning strategies, rather than depending on one particular data collection method.

\subsubsection{Effect of Group Size}

To investigate the effect of GRPO group size, we construct two variants of \methodName{} by changing the number of candidates sampled and compared in each online tuning round while keeping the same testing programs, pretrained policy, and tuning budgets. The first variant, denoted as \methodKFour, uses $K=4$ candidates per round, while the second variant, denoted as \methodKSixteen, uses $K=16$. We compare these two variants with the default \methodName{} using $K=8$ to study how the number of within-round comparisons affects online tuning performance.

\begin{figure}[htbp]
\centering
\includegraphics[width=\columnwidth]{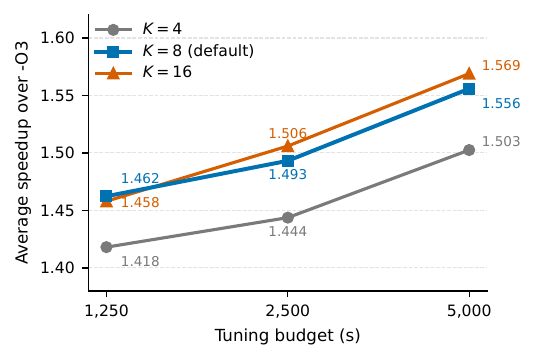}
\caption{Impact of GRPO group size on average speedup.}
\label{fig:group-size-ablation}
\end{figure}

Figure~\ref{fig:group-size-ablation} shows the average speedup under the three tuning budgets, with the exact values annotated in the figure. Overall, $K=4$ consistently performs the worst, suggesting that a small group provides too few within-round comparisons and leads to noisier advantage estimates. Increasing the group size to $K=8$ clearly improves performance across all budgets. Further increasing it to $K=16$ gives comparable performance at 1,250 seconds and only slightly higher speedups under larger budgets, indicating diminishing returns as the group grows. Since $K=16$ doubles the per-round evaluation cost compared with $K=8$, we use $K=8$ as the default setting in the main experiments, which balances advantage-estimate quality and evaluation efficiency.

Overall, the ablations support the main design of \methodName{}. Offline pretraining and program representation give the online stage a program-specific starting point, while the optimizer and group-size studies show how different online update choices affect performance under the same tuning budgets. The data-source study further shows that the same framework can work with records collected by different tuning strategies.

\section{Threats to Validity} 

\textbf{The internal threat} mainly comes from the implementation of compiler auto-tuning techniques. To reduce this threat, we utilize the reproducible packages of the compared approaches and strictly re-implement these approaches based on their corresponding publications when reproducible packages are unavailable. In addition, the authors review the implementations to further reduce this threat.

\textbf{The external threat} mainly comes from the selected benchmarks, compiler, and hardware platform. To reduce this threat, we evaluate \methodName{} on cBench and PolyBench, which are commonly used in compiler auto-tuning works~\cite{pdcat,srtuner,grouptuner,cfsca}. 
We use GCC 15.2.0, the latest GCC release available at the time of our experiments, on an x86 server platform.

\textbf{The construct threat} mainly comes from the performance metrics. Following previous compiler auto-tuning works~\cite{pdcat,srtuner,grouptuner,cfsca}, we use speedup over GCC \verb|-O3| as the evaluation metric.
Although GCC provides several optimization settings, including \verb|-O1|, \verb|-O2|, \verb|-O3|, and \verb|-Ofast|, \verb|-Ofast| extends \verb|-O3| by incorporating some unconventional optimizations that may violate certain international standards. In contrast, \verb|-O1| and \verb|-O2| do not activate many optimization flags. Consequently, \verb|-Ofast| is generally not recommended in the literature, and studies on compiler tuning are typically compared against \verb|-O3|.

\section{Discussion}

\textbf{Applicability beyond GCC flag selection.}
This work focuses on GCC optimization flag selection, but the proposed framework is not tied to GCC-specific mechanisms. It only requires an optimization interface with a well-defined action space and measurable feedback for each program. Under this abstraction, the same offline-to-online idea can be applied to other compilers and optimization tasks by redefining the actions and rewards. For example, LLVM pass ordering can use pass sequences as actions, inlining decisions can use optimization decisions as actions, and multi-objective tuning can define rewards based on execution time, code size, or their weighted combination. Exploring these extensions is left to future work.

\textbf{Combination with existing tuning techniques.}
\methodName{} is designed as an offline-to-online framework rather than a method tied to a specific search algorithm. Existing compiler auto-tuning techniques can provide historical tuning records for offline pretraining, as long as the records contain evaluated configurations and their measured performance. The data-source study in Section~\ref{sec:experimental-results} shows that records collected by different tuning strategies can be converted into an online adaptable policy. This suggests that \methodName{} can reuse search experience accumulated by existing tuners and turn it into a program-specific starting point for further online refinement.

\section{Related Work} 

Existing work on compiler auto-tuning spans both optimization ordering and optimization selection~\cite{compiler_autotuning_survey}. To position this paper, we first briefly review representative phase-ordering studies and then focus on prior work most relevant to our setting, namely flag-selection techniques.

For phase ordering, MiCOMP uses optimization sub-sequences and machine learning to predict effective optimization orders~\cite{ashouri2017micomp}. AutoPhase applies deep reinforcement learning to compiler phase ordering for HLS~\cite{autophase}. Another machine-learning-based technique predicts method-specific optimization orders in JIT compilers~\cite{kulkarni2012mitigating}. These studies focus on optimization ordering, whereas this paper focuses on flag selection.

Compiler flag selection techniques mainly differ in how they explore or restrict the optimization space. Early work treats the compiler as a black-box evaluator and searches over measured configurations: random iterative optimization repeatedly samples configurations and keeps those that improve performance~\cite{rio}, genetic and evolutionary methods evolve a population of configurations through selection and mutation~\cite{ga_gcc_options,garciarena_ga_flags}, irace uses racing to discard weak configurations under a limited budget~\cite{irace_gcc}, and OpenTuner coordinates multiple search strategies in a common tuning framework~\cite{opentuner}. Later work uses program features, historical measurements, or probabilistic models to guide configuration selection. Milepost GCC extracts program features and uses learned models to select optimization settings~\cite{milepost_gcc}; COBAYN builds Bayesian networks over program features and compiler optimizations to sample promising configurations~\cite{cobayn}; BOCA uses Bayesian optimization to select configurations under limited measurements~\cite{boca}; and CompTuner trains performance models in multiple phases and applies particle swarm optimization to search for effective flag combinations~\cite{multiphase_learning}. More recent techniques incorporate stronger structural guidance. CFSCA identifies critical flags and searches the reduced space more intensively~\cite{cfsca}; PDCAT derives constraints and preferences from compiler documentation to bias configuration generation~\cite{pdcat}; SRTuner mines synergistic relations among flags to guide compatible flag combinations~\cite{srtuner}; and GroupTuner organizes related options into groups and performs localized mutation around promising configurations~\cite{grouptuner}. Related learning-based work has also been studied in broader compiler optimization settings: CompilerGym provides reinforcement-learning environments for compiler optimization~\cite{compilergym}, MLGO trains learned policies for production LLVM decisions such as inlining and register allocation~\cite{mlgo}, and recent LLM-based approaches use large language models to predict, generate, or refine compiler optimization choices~\cite{llm_compiler_optimization,llm_compiler_foundation,decos,llm_veriopt}.

\section{Conclusion}

To alleviate human efforts on compiler tuning for a specific program, in the literature, several compiler auto-tuning techniques have been proposed, but all suffer from performance issues. In this paper, we proposed \methodName{}, an offline-to-online reinforcement learning framework for compiler flag auto-tuning. \methodName{} first learns a program-conditioned policy from historical tuning records, and then refines this policy on the target program using real compile-run feedback. In our implementation, GRPO serves as the default online optimizer by comparing groups of candidate configurations evaluated for the same program. According to the experimental results, \methodName{} achieves the best overall runtime performance among the compared compiler auto-tuning techniques. Moreover, the ablation analysis demonstrates the contribution of the main components in \methodName{}.

\section{Data Availability}
We make the source code and data of \methodName{} available on website~\cite{zenodo}.

\bibliographystyle{IEEEtran}
\bibliography{references}

\end{document}